# LDPC Codes in Cooperative Communication

**Ali Mehrban**

*a.mehrban@ieee.org*

**Supervised by:**

Professor Martin Johnston

*martin.johnston@ncl.ac.uk*

## ACKNOWLEDGEMENT

In the name of greatest All mighty God who has always bless me with potential knowledge and success.

I am thankful to my supervisor Professor Martin Johnston, my friends who help me during my hard times when I need their assistance during simulation.

I am especially thankful to my Parents, who had always provided me the courage, strength, best wishes, moral and financial support during my study.

# Abstract

In fact, the broadcast nature of every transmitter makes it possible for other transceivers in the channel to overhear the broadcasted signal. The proposed idea in cooperative communication is to use these intermediate transceivers as relay for the transmitted signal, therefore we will have spatial diversity which can improve throughput and received data reliability in the system. In this dissertation we consider some important aspects of cooperative communication in a network composed of three nodes. First, we verify the increase of reliability for the received signal by comparing the reliability of the received bits in a cooperative network with a non-cooperative one. Then we step forward and use LDPC error correction technique to improve the reliability of the received bits even more and compare it with a network without LDPC codes (encoder and decoder) to measure the level of improvement for different SNRs. The overall aim of this dissertation is to deploy cooperative communication idea to consider and test its claimed aspects and also enhance its performance by using LDPC error correction technique.

# LIST OF CONTENTS

# Chapter 1
Introduction

## 1.1 Background

In wireless communications in terms of signal path there is no specified constraint for a transmitted signal through a channel, in other words in wireless communications the nature of transmitted signal is broadcast. Also according to the nature of a wireless channel there could be many different types of objects in the way of a transmitted signal such as buildings, towers and so on which each one of them can cause some affects for the broadcast signal like diffraction, refraction, reflection and scattering, it means in fact just one signal is transmitted but many different samples of the same signal are received in the destination (receiver). Thus, received signals have different path lengths which result in received signals with varying phase and amplitude. The amplitude fluctuation of a signal is called fading (multipath fading) which makes it hard for the receiver to decide on the received signal. The solution to overcome this problem is called "diversity gain" which means to feed the receiver with different samples of the same signal therefore we will be able to use various techniques such as MRC (Maximum Ratio Combining) to ease the way for the receiver to extract right ones and zeros from received values.

To attain this, we can use frequency, space or time resources which are named frequency, space and time diversity respectively. Multiple antennas in transmitter and receiver are used to provide space diversity and in this way the receiver will be fed with multiple copies of the transmitted signal and this type of communication is named Multiple Input Multiple Output (MIMO) communications but when multiple antennas are only used in receiver or transmitter this is called receive or transmit diversity. But there will be a dilemma meanwhile and that is, in various types of wireless communications like mobile communications or Ad hoc networks the devices used as mobile nodes of a network are quite small devices which it is not feasible to embed multiple antennas in them and here is the time which cooperative communication is proposed as a solution to provide space diversity for such wireless networks. It was first proposed by Sendonaris et al [10]. As a definition cooperative communication is a way in which each node of the network not only generates and transmits its own data but also acts as an assistant for the other nodes. This means it receives the signals of other devices in the network and amplifies and forwards them to the destination. Therefore, in this case this node is called a "relay". Hence it is obvious that the number of copies depends on the number of the assistant nodes. At first look it looks like the power usage of each cooperating user has increased because not only it sends its own data but also forwards the other users' data to their meant destinations. But according to research carried out in this case it has been proved that the overall transmit power of the system is reduced. Cooperative communication is a subgroup of relay channels but differs in one thing that the relay node not only relays the received signal but also has its own data to send.

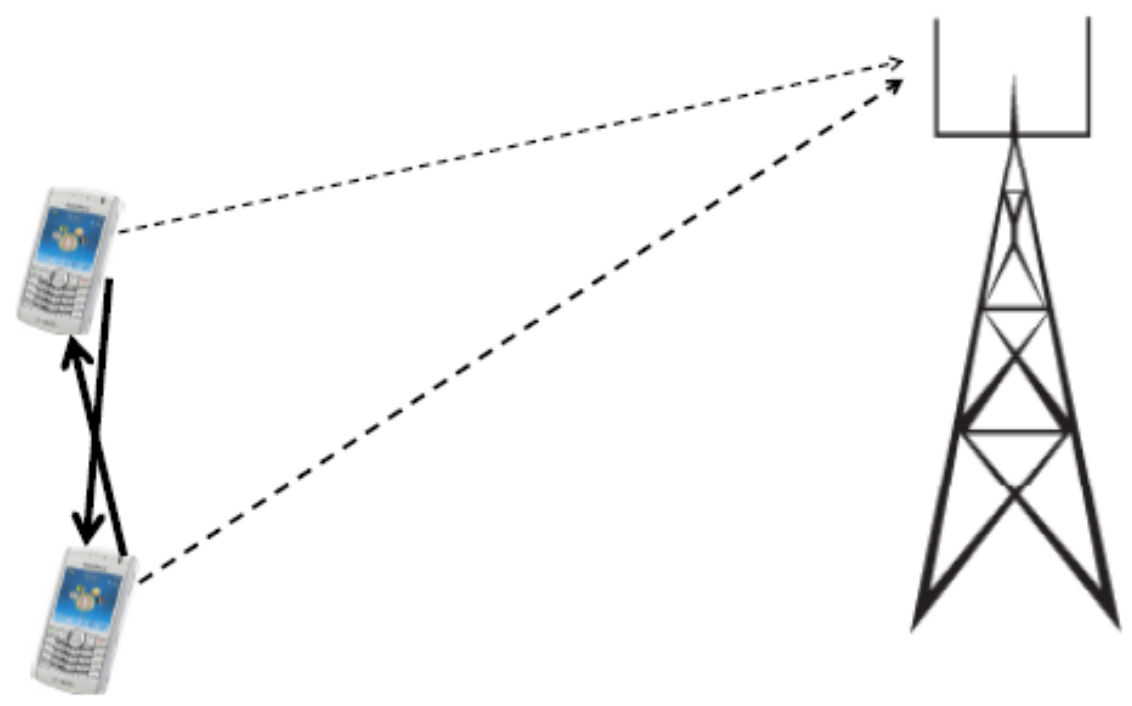

Fig. 1. cooperative communication setup

Relay channel first was proposed by Van Der Meulen [25] and then was developed by El Gamal [8]. They considered different types of relay channels and discussed the capacity of each one of them. Eventually their attempts formed some relaying protocols as listed below which will be elaborated in next chapter:

1) *Amplify-and-forward*
2) *Decode-and-forward*

## 1.2 Aims/Objectives

This dissertation is divided into five chapters as mentioned below:

**Chapter One : Introduction**

**Chapter Two: Cooperative Communication in a 3-node Network (without LDPC coding)** In this chapter we explain wireless channels specially additive white Gaussian noise channels and then we will go forward for Characterizing performance through channel capacity and then we will come to cooperative communications in detail by starting from relay channels and protocols specially for cooperative communications with single relay. The next topic which comes in this chapter is the modulation scheme that has been used in this project and then the optimal way of combining received signals is the next topic. The chapter will be continued with some simulations that have been done for a 3- node network without use of LDPC coding to compare the performance of a wireless cooperative network with non-cooperative one. And eventually the chapter will end up with an explanation about some applications of cooperative communications.

**Chapter Three: LDPC Coding**

In this chapter we will elaborate LDPC coding from the emergence until now and will go in details from coding to decoding and even will provide an explanation for decoding steps and at the end we will consider its performance and compare it with different error correction coding techniques such as turbo codes.

**Chapter Four: LDPC Coding in Cooperative Communications**

In this chapter we will come back to chapter one simulation part and develop it with use of LDPC coding and then make a comparison of performance of the network in the existence and absence of LDPC coding to gain more knowledge about the improving effects of LDPC coding on a system performance.

**Chapter Five: Conclusion**

In this chapter we summarize the whole project efforts and make a short conclusion by consideration of the previously elaborated results.

Then we will list the References that have been used for this project from IEEE papers to books and PhD thesis.

# Chapter 2

# Cooperative Communication In a 3-Node Network

(Without LDPC Coding)

## 2.1 Wireless Channels

Nowadays wireless communication is facing fast technological advances in terms of devices deployed in wireless networks and architectures of wireless networks. These advances have led wireless communication to see beyond a simple point to point communication, in other words as a result of broadcast nature of wireless signal transmission which can be overheard by other users, it has converted the users to an active part of the network which can receive and process the other users' data rather than being just a simple end user. This can be counted as the advent of cooperation in networking which yields higher performance gain.

By technological progress in fields of energy storage, antennas and integrated circuits, digital signal processors, etc. all make it feasible in practice to be implementable. In fact, the main reason of resorting to this technology is gaining the same advantages of MIMO systems. Those advantages are improvements in performance, data communication speed, communication capacity (faster transmission-higher bit rate), battery consumption (will be reduced), network lifetime extension, transmission coverage expansion, higher throughput, etc.

The notable point in wireless communications is wireless channel, because the different conditions of a medium in different ranges from one device to the other one will end up with impairment to the signal. These effects can be exhibited in the shape of noise, interference, distortion and attenuation. Also Noise can be additive like additive white Gaussian noise (AWGN) and channels can be modeled by their attitude towards signals for example additive white Gaussian Noise channel is the simplest one in this case and can be modeled as below in which y(t) is the output and x(t) is the input of the channel and Γ is the attenuator of main signal that cause loss in power and n(t) is an additive noise that in fact is a random variable with Gaussian distribution :

$$y(t)= x(t)/\sqrt{\Gamma} + n(t) \qquad (2\text{-}1)$$

And Gaussian distribution has a curve as shown:

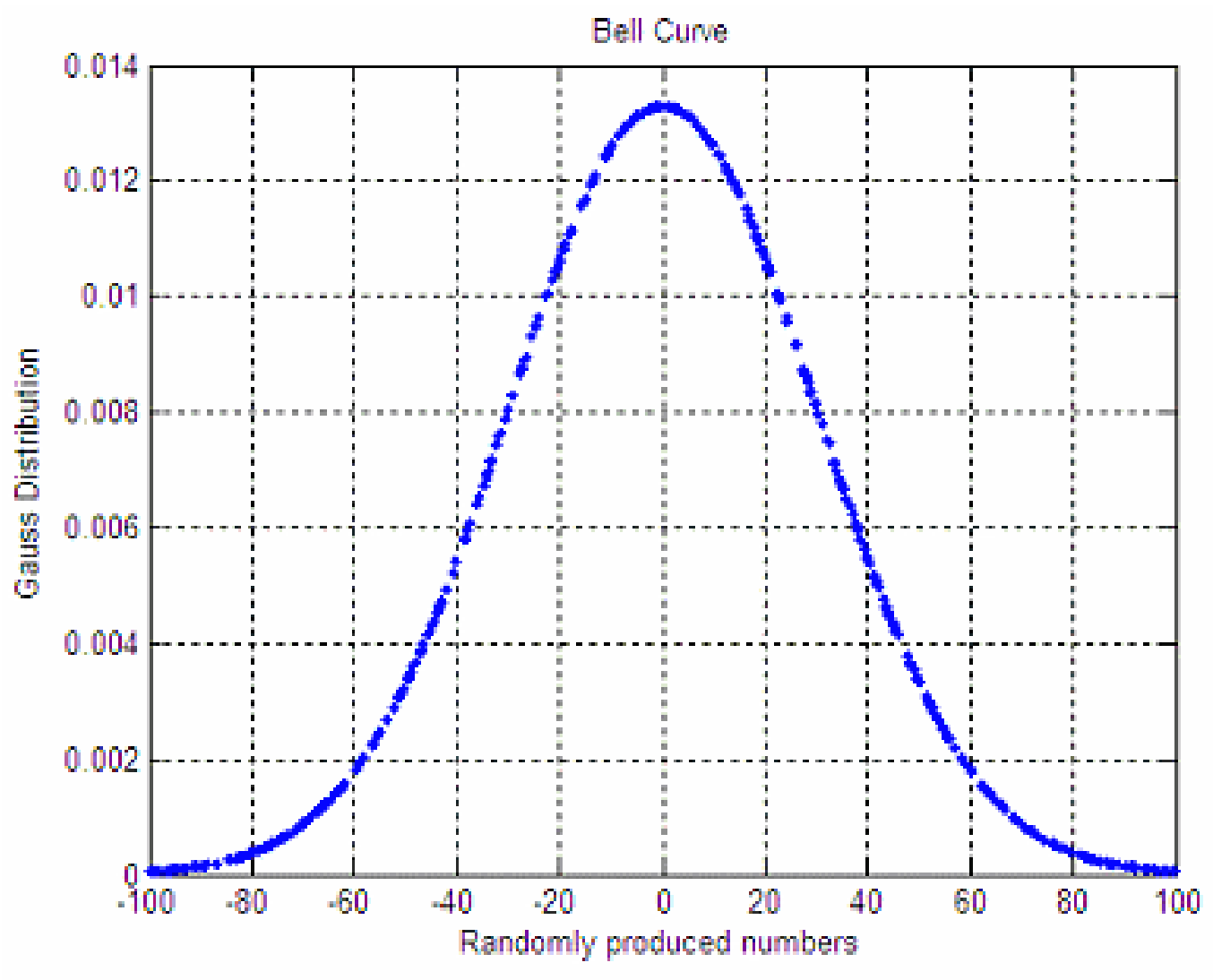

**Figure 2 Gaussian Distribution**

Whatever signal travels longer through the channel will lose its power and this is called path loss which is an important factor in power reduction and is measured by decibels (dB) of the received and transmitted signal. It is defined as below:

$$\Gamma_{dB} = 10v\,\log(d/d0) + c \qquad (2.2)$$

As seen $\Gamma_{dB}$ is the measured pass loss in decibel and *d* is the distance between receiver and transmitter and $d_0$ is the distance to a premeasured power reference point and c and v are constant values. Obviously, it is clear that path loss is dependent on the distance between receiver and transmitter antennas, but a noticeable point is if we assume a transmitter is transmitting a signal towards two antennas which are in equal distance to the transmitter antenna the power loss will vary and the reason is because signal travels through different paths and encounters difference objects on the way. This is called shadow loss (shadow fading). Therefore, it will be a random variable since the place of obstruction and the nature of them is not known to us. Since signals faces different scatterers, attenuators and reflectors during the path so it will cause the arrival of multiple copies of the signal in receiver and since their phases and amplitudes may be different thus it will cause fading in received signal and this affect is called multipath fading and channel is multipath channel. There are even more factors affecting a multipath channel such as speed of mobile devices in the network and transmission bandwidth. As a result of mobility of devices, the positions of surrounding obstructions will vary with the time and It will cause the characteristics of such channel to be absolutely random. Here I want to extract the impulse response of the channel, firstly the received signal can be shown as below:

$$y(t)= \sum_{i=1}^{L} hi\ (t)x(t\ -\ \tau i\ (t)) \qquad (2\text{-}3)$$

Here y(t) is received signal and x(t) is transmitted signal over time and hi(t) is the attenuation of the i-th path at time t, $\tau_i$ (t) is corresponding path delay and L denotes the number of existing paths. From this formula impulse response of the channel can be extracted as below:

$$h(t,\tau) = \sum_{i=1}^{L} h_i(t)\delta(t - \tau_i(t)). \qquad (2\text{-}4)$$

But in digital communications it is easier to consider the discrete time model of the channel which comes as below:

$$y[m] = \sum_{k=l}^{L} h_k[m]x[m-k] \qquad (2\text{-}5)$$

Therefore, as shown, the receiver samples the arrival signals periodically with different channel coefficients and here in this formula, $h_k$ (m) are the channel coefficients. In fact, this formula shows a digital FIR (Finite Impulse Response) filter. Signal received from each path is modeled by a corresponding coefficient and according to the nature of each path hence we can expect the values of the impulse responses $h_k$ [m] to be random. The function resulted from the average power of corresponding path is called “power delay profile” and for each wireless channel a power delay profile is proposed as standard like the one below which is a modified version of reference channel model proposed by ITU that is named “vehicular B”. Since there may be frequencies which are attenuated much more than the threshold level of sensitivity of receivers, therefore this profile is seriously dependent on the sensitivity of receivers.

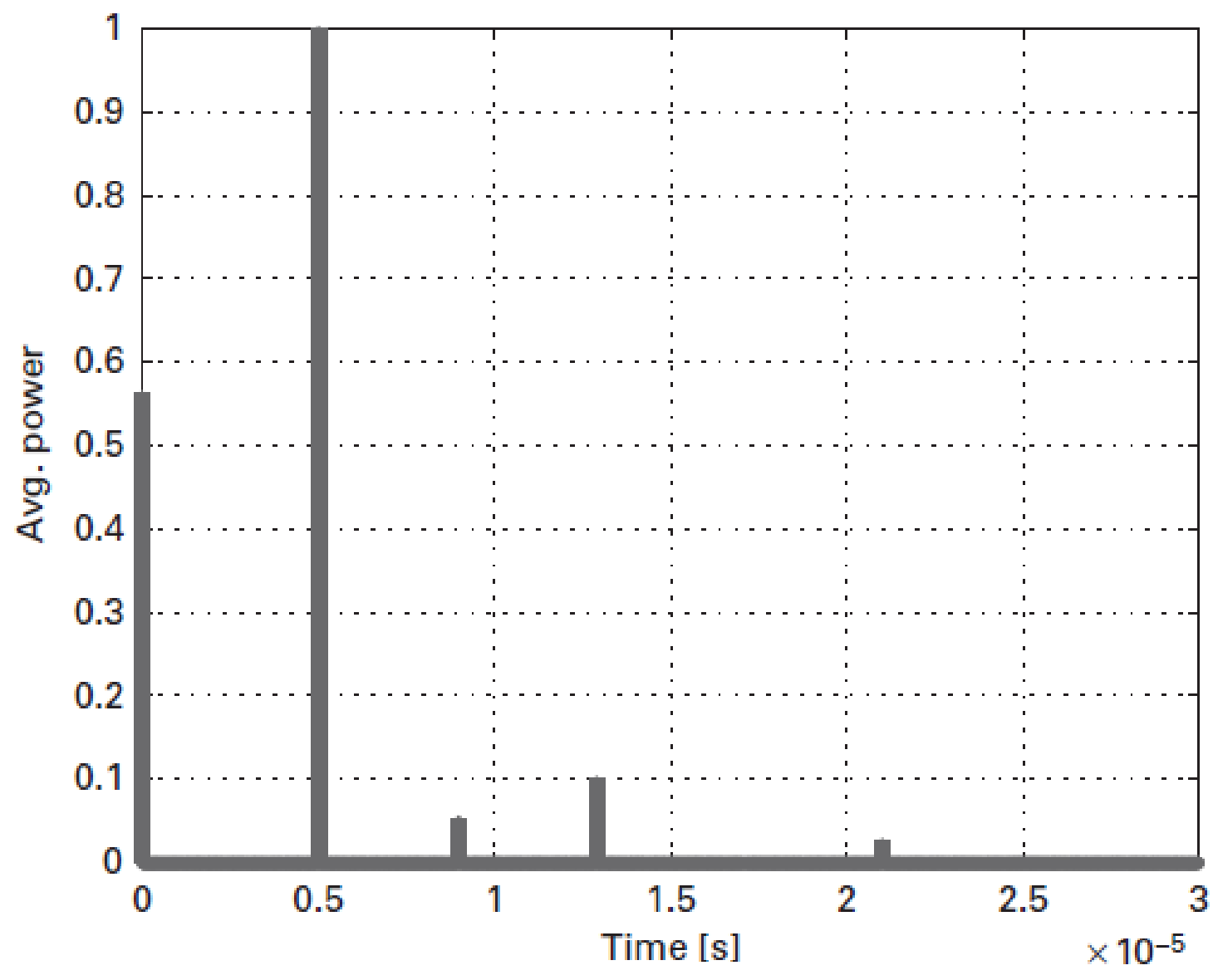

The power delay profile of a typical wireless channel.

**Figure 3**

Some parameters can be realized from this power delay profile which can be defined as below:

> “• The *channel delay spread* is the time difference between the arrival of the first measured path and the last. If the duration of the symbols used for signaling over the channel exceeds the delay spread, then the symbols will suffer from inter-symbol interference. Note that, in principle, there may be several signals arriving through so attenuated paths, which may not be measured due to sensitivity of the receiver. This makes the concept of delay spread tied to the sensitivity of the receiver.

• The *coherence bandwidth* is the range of frequencies over which the amplitude of two spectral components of the channel response is correlated. The coherence bandwidth provides a measurement of the range of frequencies over which the channel shows a flat frequency response, in the sense that all the spectral components have approximately the same amplitude and a linear change of phase. This means that if the transmitted signal bandwidth is less than the channel coherence bandwidth, then all the spectral components of the signal will be affected by the same attenuation and by a linear change of phase. In this case, the channel is said to be a *flat fading channel*. In another way, since the signal sees a channel with flat frequency response, the channel is often called a *narrowband channel*. If on the contrary, the transmitted signal bandwidth is more than the channel coherence bandwidth, then the spectral components of the signal will be affected by different attenuations. In this case, the channel is said to be a *frequency selective channel* or a *broadband channel*.

• The *Doppler spread* is defined as the range of frequencies over which the Doppler power spectrum is nonzero. The Doppler spread is the inverse of the channel coherence time and, as such, provides information on how fast the channel changes over time. Here, again, the notion of how fast the channel is changing depends also on the input signal. If the channel coherence time is larger than the transmitted signal symbol period; or equivalently, if the Doppler spread is smaller than the signal bandwidth, the channel will be changing over a period longer than the input symbol duration. In this case, the channel is said to have *slow fading*. If the converse applies, the channel is said to have *fast fading*." [P7 - Cooperative Communications and Networking [5]]

## 2.2 Channel Capacity

Determination of channel capacity is dependent on the channel changes over a period like coding interval. Therefore, while random variations of the channel are ergodic and stationary we can use the common concept of capacity introduced by Claude Shannon. Thus, for an AWGN channel when channel characteristics is known for the receiver and there is flat fading in the channel the formula can be as below:

$$C = \mathrm{E}\left[ \log\left(1 + \frac{|h|^2 P}{N_0}\right)\right] \quad (2\text{-}6)$$

Here $N_0$ is the variance of noise and $|h|^2$ is the envelope of channel attenuation and P is the power of transmitted signal. If channel has non-ergodic variations with Rayleigh distribution its Shannon capacity will be very small or even zero which resulted from deep fades. Therefore, in this case a new concept is proposed which is called outage capacity. More precisely the outage capacity is defined as “the set of channel realizations that cannot support reliable transmission at a rate R. In other words the outage event is the set of channel realizations with an associated capacity less than a transmit rate R ” [P25- Cooperative Communications and Networking[5]]. Hence the outage condition comes as below:

$$\log\left(1 + \frac{|h|^2 P}{N_0}\right) < R. \quad (2\text{-}7)$$

Then the outage probability can be calculated as below:

$$\mathrm{P_{out}} = \Pr\left[ \log\left(1 + \frac{|h|^2 P}{N_0}\right) < R\right] \quad (2\text{-}8)$$

Then outage capacity is written as below:

$$\Pr\left[C \le C_{\mathrm{out}}\right] = \Pr_{\mathrm{out}} \quad (2\text{-}9)$$

$C_{out}$ is a sort of threshold for information rate that can be transmitted and received reliably through the channel and C is Shannon capacity of the channel. Outage probability can be defined in another way via signal to noise ratio. Here a threshold for SNR is assumed then the formula can be written as below:

$$\mathrm{P_{out}} = \Pr\left[\gamma < 2^R - 1\right] = \Pr[\gamma < \gamma_{\mathrm{T}}] \quad (2\text{-}10)$$

Here $\gamma$ is signal to noise ratio (SNR) and $\gamma_T$ is threshold SNR.

## 2.3 Cooperative Communications

Cooperative communication was first introduced to provide the benefits of MIMO systems for single antenna mobiles. Wireless cooperative communication is related to a wireless network such as cellular or ad hoc networks and to enhance the reliability and data rate of a system without increasing power or bandwidth, therefore, to realize the advantages of cooperative communications, firstly a brief look at MIMO systems seems essential. MIMO is an antenna technology which uses multiple antennas in both transmitter and receiver devices with various configurations. MIMO technology improves channel capacity, quality and coverage of the network, in other words it improves system capacity, range and reliability via using space dimensions. In MIMO systems the relationship between number of antenna elements and channel capacity is linear while in SIMO/MISO systems it is logarithmic. And obviously as much the number of antennas deployed in transmitter and receiver is increased the channel capacity will be increased too. In simple words MIMO takes advantage of multipath to provide multiple samples of a signal for receiver and uses an algorithm to process the received signals in order to extract the original transmitted signal out of it. The image below shows the structure of a MIMO system.

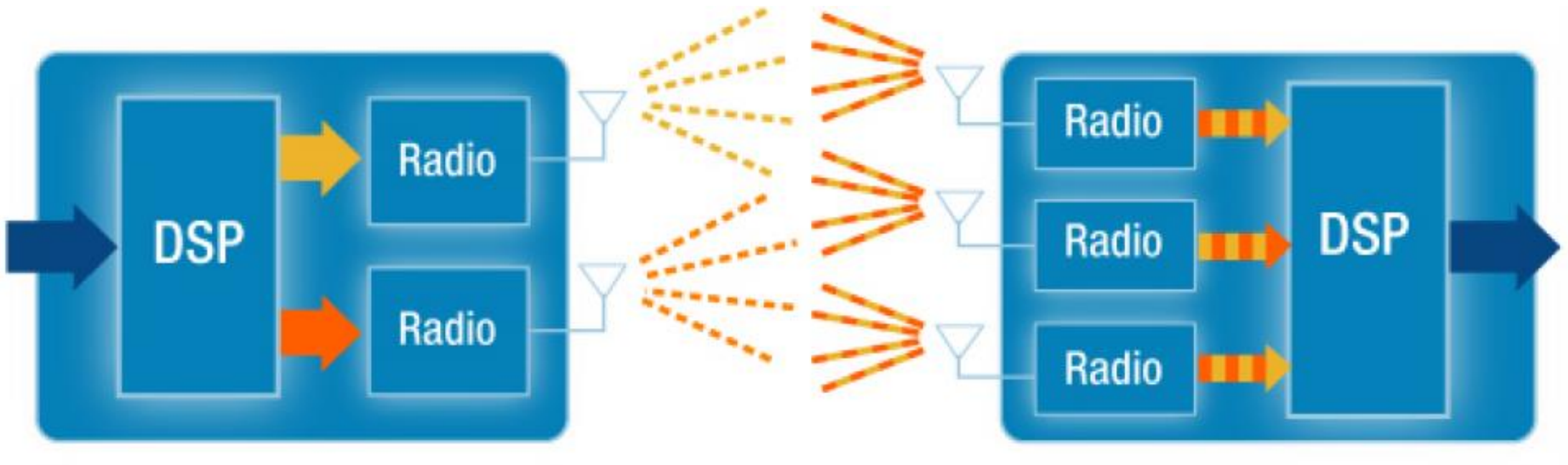

**Figure 4 MIMO System**

MIMO applications are widespread from WIMAX 802.16 e and WIFI 802.11 n to 4G and RFID. As previously explained the idea of cooperative communication is to provide the same advantages of MIMO systems for single antenna wireless systems with this important distinction that instead of using multiple antennas in transmitter, uses other neighboring users of the network to provide the receiver with multiple copies of the signal.

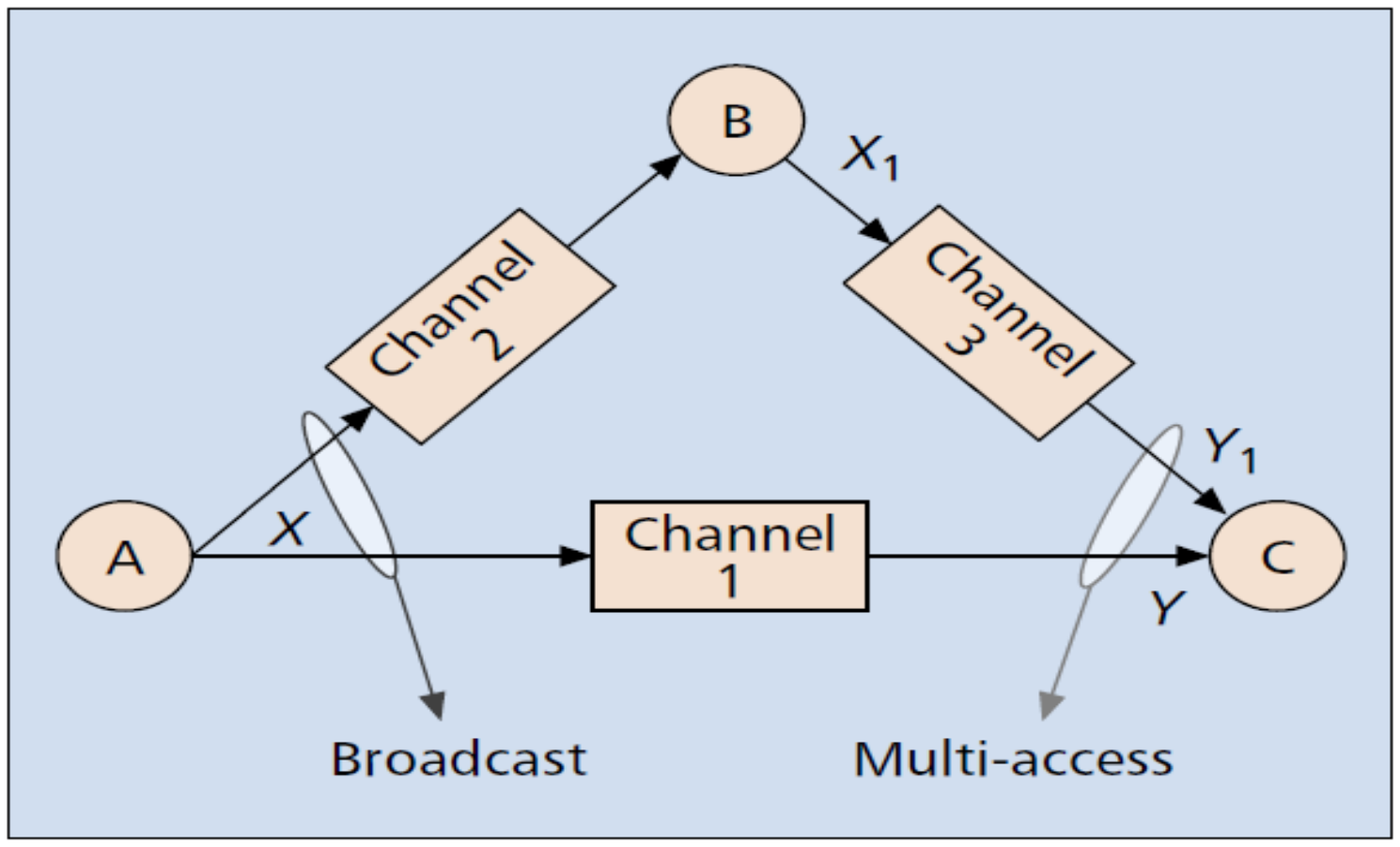

**Figure 5   Single Relay Cooperative Communication**

In this dissertation we only consider a single relay system. In a single relay cooperative network, a 2-phase strategy has been used to avoid interference between phases. The first phase is transmission of information from the source which will be received by destination and partner at the same time and then in the second phase the partner retransmits the information to destination to facilitate making a correct decision over the received information for the receiver.

### 2.3.1 Cooperation Protocols

A) **Amplify and Forward (AF):** In this protocol like destination the relay user receives the noisy version of the signal and then amplifies and forwards it to the destination. Destination also combines the two received noisy faded signal, hence can make a better decision on information. This method was proposed by Laneman [2]. It is important to know in this protocol to do optimal decoding it is assumed that the base station has knowledge about the inter-user channel coefficients.

B) **Decode and Forward (DF):** In this protocol every cooperating user receives the noisy version of the signal like in AF, but the difference is that here the information in relay node is decoded and then re-encoded and retransmitted to the destination. The main advantage of this protocol over AF is that in DF at the relay the effects of additive noise are cancelled or reduced.

**Relay Modes:**

**Full Duplex:** Relay can receive and transmit at the same frequency band and at the same time.

**Half Duplex:** Relay cannot receive and transmit at the same time and the same frequency band.

Now in this dissertation we will have a discussion on AF protocol and a simulation for experimenting the improving effects of cooperative communication by making a comparison between a cooperative and a non-cooperative communication performance.

First it is assumed that there is a network composed of three nodes: transmitter, relay and destination. The relay is assumed to be full duplex so we can expect simultaneous transmission and detection in the system. The channels between source and destination or relay and destination are called uplink channels and the channel between source and relay is called inter-user channel. Here in this dissertation we consider the signal in two time slots and the uplink channels are assumed to be Rayleigh flat fading channels and also all channels are one-sided it means there is no signal coming back to source. Assume in first time slot $x_s$ [n] is transmitted by the source therefore the signals reaching relay and destination are calculated as below:

$$y_{sr}[n] = \sqrt{\varepsilon}\ a_{sr}x_s[n] + z_{sr}[n]$$
$$y_{sd}[n] = \sqrt{\varepsilon}\ a_{sd}x_s[n] + z_{sd}[n] \qquad \text{(2-11), (2-12)}$$

In this formula $\varepsilon$ is the power of transmitted signal and $y_{sr}$ *[n]* and $y_{sd}$*[n]* are received signals at the relay and destination respectively. Also the terms $Z_{sr}$*[n]* and $Z_{sd}$*[n]* denote the additive white Gaussian noise with zero-mean variances $N0_{sd}$ *and* $N0_{sr}$. $a_{sr}$ is fading coefficient of inter-user channel (with variance $\sigma_{sd}^2$ ) that here is assumed to be fixed. $a_{sd}$ and $a_{rd}$ are also the coefficients of Rayleigh flat fading channels with variances $\sigma_{sd}^2$, $\sigma_{rd}^2$ which end in destination node.

In amplify and forward protocol to overcome the fading effects of the channel between relay and source, the signal is amplified in relay and then will be forwarded to the destination. Thus, amplifying factor is calculated via this formula:

$$y_{rd}'[n] = \sqrt{\varepsilon}\, a_{rd} x_r'[n] + z_{rd}'[n] \quad (2\text{-}13) \qquad \beta = \sqrt{\frac{1}{|a_{sr}|^2 \varepsilon + N0_{sr}}} \quad (2\text{-}14)$$

Therefore, the output of the relay node will be as below:

$$x_r'[n] = \beta\, y_{sr}[n] \quad (2\text{-}15)$$

Also channel signal to noise ratio (SNR) is defined as:

$$\gamma_{ij} \equiv |a_{ij}|^2 \varepsilon / N0_{ij} \quad (i=s,r,\ j= r,d) \quad (2\text{-}16)$$

And going even further the expected value of the channel SNR is calculated from this formula:

$$\overline{\gamma_{ij}} \overset{\Delta}{=} E[\gamma_{ij}] = E[|a_{ij}|^2]\varepsilon / N0_{ij} = \sigma_{ij}{}^2 \varepsilon / N0_{ij} \quad (2\text{-}17)$$

After two time slots the received signals at the receiver should be combined, also there are various techniques to combine these two received signals but the most optimized technique which maximize the overall signal to noise ratio is the maximal ratio combiner (MRC). In this technique a coherent detector that knows the channel coefficients is required. Below is the structure of an optimal combiner at the destination:

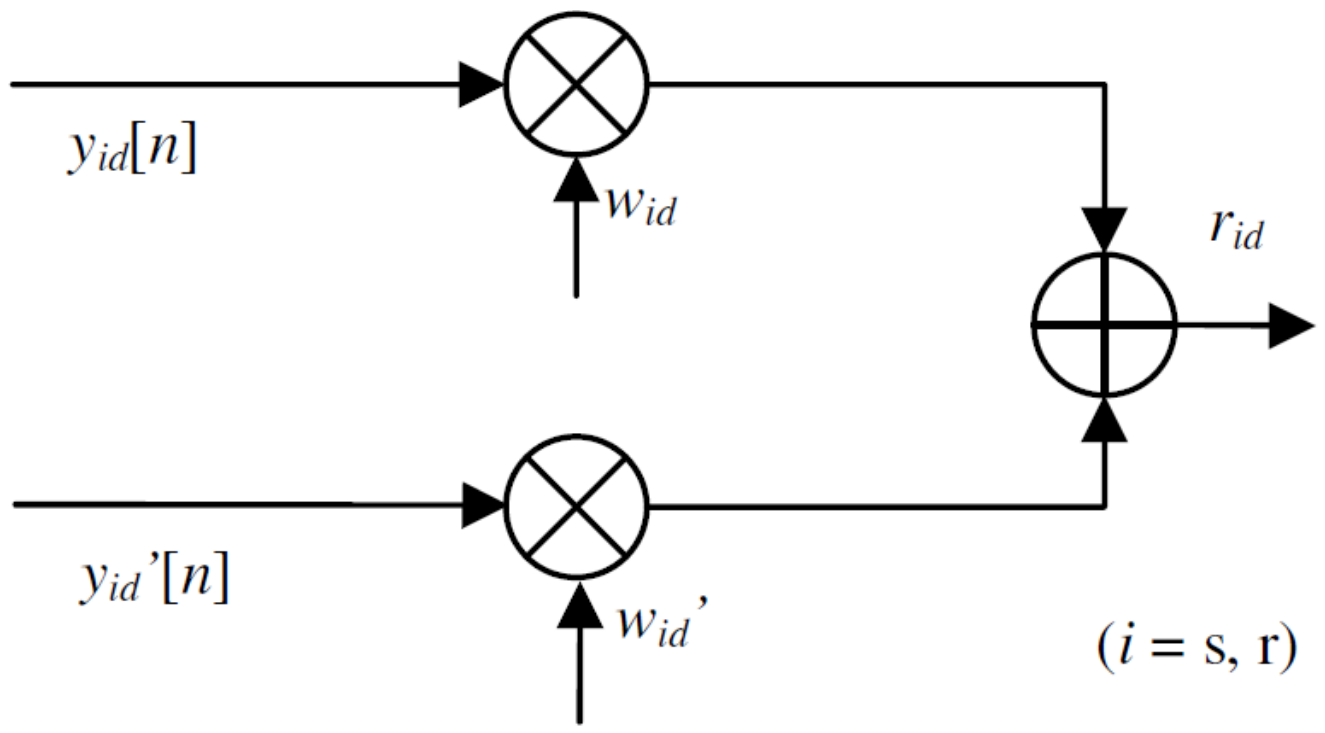

**Figure 6 Optimal Combiner at the Receiver**

$w_{id}$ and $w_{id}'$ are optimal combining gains and to design an acceptable performing combing gains the channel knowledge is assumed to be available at the receiver. The output of this optimal combiner will be calculated as:

$$r_{sd}[n] = w_{sd}y_{sd}[n] + w_{rd}'y_{rd}'[n] \qquad (2\text{-}18)$$

Combining gains also can be calculated via these two formulas:

$$w_{sd} = \frac{a_{sd}^{*}\sqrt{\varepsilon}}{N0_{sd}}, \quad w_{rd}' = \frac{a_{rd}^{*}\beta a_{sr}^{*}\sqrt{\varepsilon}}{|a_{rd}|^{2}\ \beta^{2}N0_{sr} + N0_{rd}} \qquad (2\text{-}19), (2\text{-}20)$$

After it all the output of combiner ($r_{sd}$) goes through a decoder to give us the estimated received values for source main message.

### 2.3.2 Simulation of Single Relay Cooperation Network

In this simulation we have a cooperative network composed of three nodes (single relay) and the relaying protocol used in this simulation is amplify and forward method. The network is a one-sided full duplex network that has been made up of 3channels: Source – Destination, Source – Relay and Relay – Destination. Here it is assumed that the Source-Relay channel is an additive white Gaussian noise channel with a fixed SNR depending on different simulations sometimes 5 dB and sometimes 10 dB. But the Source- Destination and Relay – Destination is assumed to be Rayleigh fading channels with varying signal to noise ratio from 0 to 15 dB. So according to the formula below $N0_{ij}$ Can be easily achieved:

$$N0_{ij} \equiv |a_{ij}|^2 \varepsilon / \gamma_{ij} \qquad (i=s,r,\ j=r,d) \qquad (2\text{-}21)$$

## 2.3.2.1 BPSK Modulation

In this simulation no error correction coding is used and the modulation scheme used is binary phase shift keying (BPSK) which is the simplest type of phase shift keying modulations. It has two phases with 180$^o$ phase difference. From performance point of view, it is the most robust modulation scheme among PSKs because it can withstand a higher level of noise in compare to the other schemes, in other words in order to cause the BPSK demodulator to make a wrong decision in compare to the other PSK schemes a higher level of noise is needed. Therefore, it can be a reliable modulation and in fact it is the main advantage of this modulation scheme. But this scheme, in comparison to the other PSKs has some disadvantages too and the main disadvantage is its low data rate. Simpler to say BPSK can modulate only one bit per symbol (1bit/symbol), and it means this is not an appropriate modulation scheme for high date rate communication when the bandwidth is limited. The position of the constellation points does not matter, and in overall the phases are assumed 0 and 180$^o$ and the conveyor signals can be obtained by solving the following equation:

$$s_n(t) = \sqrt{\frac{2E_b}{T_b}}\cos(2\pi f_c t + \pi(1-n)), n = 0, 1. \qquad (2\text{-}22)$$

By solving the following equations, the result will be two conveyor signals written as below:

$$s_0(t) = \sqrt{\frac{2E_b}{T_b}}\cos(2\pi f_c t + \pi) = -\sqrt{\frac{2E_b}{T_b}}\cos(2\pi f_c t) \quad (2\text{-}23)$$

$$s_1(t) = \sqrt{\frac{2E_b}{T_b}}\cos(2\pi f_c t) \quad (2\text{-}24)$$

$s_0$ is assigned for binary "one" and $s_1$ is assigned for binary "zero" and $f_c$ is the carrier frequency.

#### 2.3.2.2 Bit Error Rate (BER)

In digital communication several bits are transmitted, and several bits are received, therefore because of the existence of interference, noise and distortion the received stream may be different from the transmitted one in some bits, this difference is called bit error rate which is always expressed as percentage. Also, the probability of bit error rate occurrence is called bit error probability. In a communication system the bit error rates for different signal to noise ratios are different and the result can be drawn as a BER/SNR graph to test the performance of the system.

Overall, as much for a defined ranges of SNRs the BER values are smaller it means the performance of the system is better and in other words the system is more reliable because noise and interference across the channel has less effects on the system reliability. That is why to experiment the performance of a cooperative network we use BER/SNR graph and compare it with the BER/SNR curve of a non-cooperative system to prove the capabilities of a cooperative network. This is noticeable that whatever the number of transmitted data frames are the reliability of this graph will be higher and that is why we use one thousand frames for the following simulation.

Thus, in this simulation the output is two resulting BER/SNR curves for cooperative and non-cooperative and via analyzing the curve we can realize the effect of cooperative communication. The resulted curves are shown in the figure below:

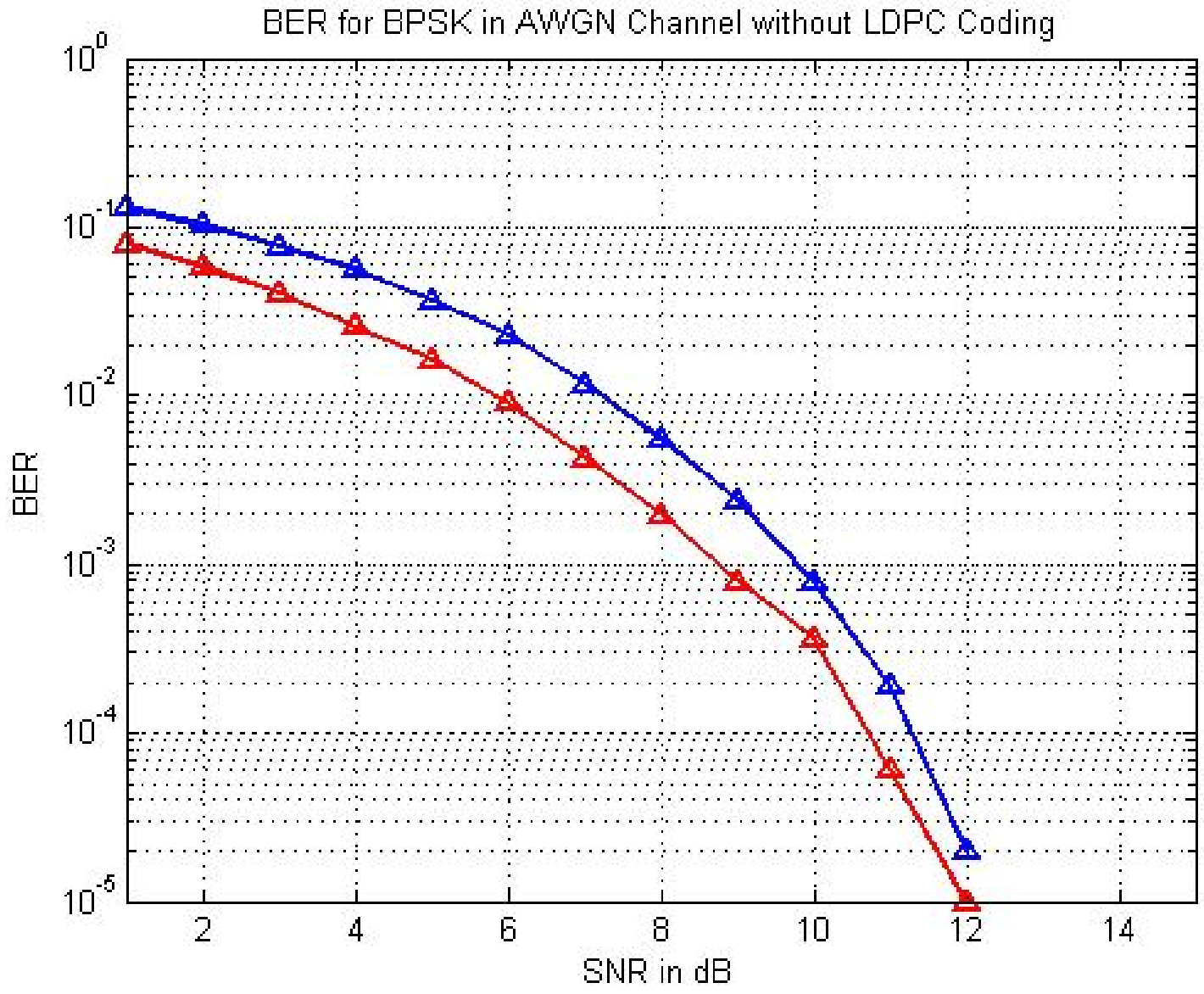

**Figure 7 Cooperative vs Non-Cooperative** ( ---- **Cooperative** ---- Non-Cooperative)

As it is shown the red curve shows bit error rates of a cooperative network versus bit error rates of a non-cooperative network (shown by blue line) over a range of SNRs from zero to 15 dB. As previously said the SNR of the inter-user channel must be fixed and here in this simulation it is assumed equal to 5 dB. The fading coefficients of relay channels are determined by the following formulas to simulate the relay channel characteristics.

$$a_{sd}(k) = \sqrt{x^2(k) + y^2(k)} \qquad (2-25)$$

$$a_{rd}(k) = \sqrt{x^2(k) + y^2(k)} \qquad (2-26)$$

Here x, y are random numbers resulted from the summation of 0.5 and a random number between 1 and 0. Ks are incrementing numbers from 1 to 15 according to the varying length of Source – Destination channel. Also, the fading coefficient of inter-user channel which is not a fading Rayleigh channel is assumed to be one. The length of message is two hundred bits, and the number of data frames used for this simulation is one thousand frames. As expected earlier it shows despite not using any error correction technique the performance of a cooperative communication system is better than a non-cooperative one because it has less bit error rates for the same SNRs, and this is exactly what was aimed from proposing a cooperative communication idea. As much the SNR of inter-user channel is higher (in dB) the bit error rate performance of the system becomes better. Below is a figure showing the BER/SNR curves of a cooperative network with different inter-user channel SNRs.

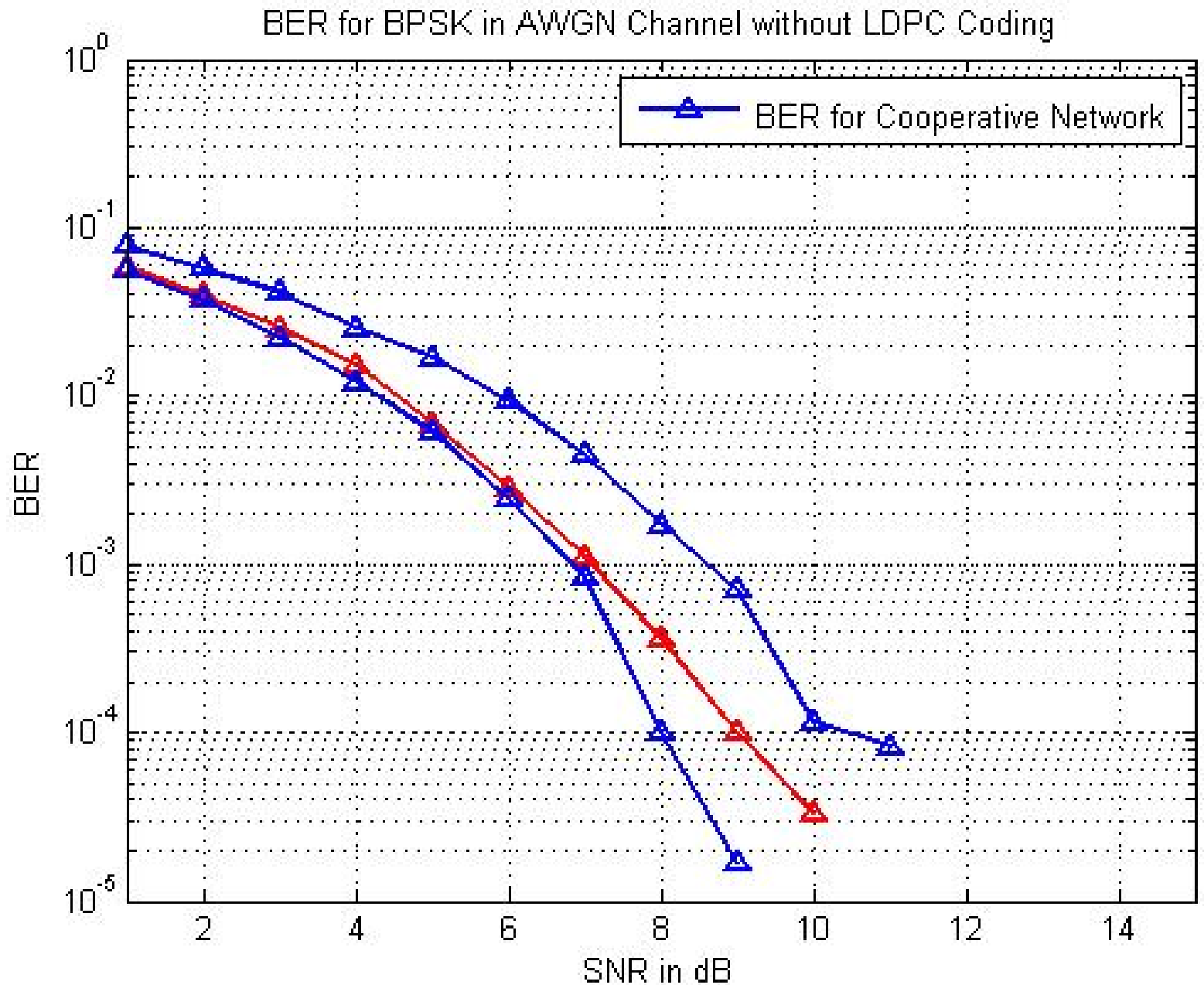

**Figure 8 Cooperative (Inter-user Channel SNRs= 5, 15, 30)**

In graph above there are three curves for respectively from the top to the bottom SNR= 5,15,30 dB. And as it is easily understood that whatever the SNR of inter-user channel is becoming higher the Curve is tending to the bottom more and this indicates a better performance of the system.

## 2.3.3 Applications

As previously has been mentioned while our communication devices are becoming smaller, using transmission diversity is the best way to provide more functions and improve the system performance. In addition to all of them, transmit diversity makes our high-speed communications more reliable. Hence the main application of cooperative communication is in the mobile industry where its movement is towards fourth generation of mobile communications (LTE - Advanced). To satisfy greedy demands on multi–Giga bit per second data rates, use of this sort of communication in LTE-Advanced will be standardized. Also, cooperative communication has found its way in other fields such as military communications and disaster management by forming an ad hoc network in which each user relays the received data of its partners to different meant destinations. Also, cooperative communication has found its way in wireless sensor networks to reduce the overall energy consumption of the network. In future the major vision of communication technology is to look beyond a simple point to point communication, hence the proliferating use of networks such as ad hoc networks that also use cooperative communication is inevitable. Therefore, here will have a brief discussion about wireless ad hoc networks as a type of cooperative communication.

### 2.3.3.1 Ad Hoc

In fact, a wireless ad hoc network without any center means an ad hoc network is not reliant on routers or access points as a predefined structure. Moreover, a wireless ad hoc network has a fluid shape, it means because of the mobility of wireless devices in the network the network structure is not fixed. In a wireless ad hoc network, all nodes participate in routing and forwarding data. And based on network connectivity, it is specified that which node's turn is to route the data or better to say which node is in a better circumstance to route the data. The first shape of wireless ad hoc networks were packet radio networks which emerged in the continuation of ALOHAnet project directed by DARPA in 1970s. After that more wireless ad hoc network were introduces that all of them can be classified in three major applications:

1) Mobile ad hoc networks (MANET)
2) Wireless mesh networks (WMN)
3) Wireless sensor networks (WSN)

Since mobile and ad hoc network is more relevant to our aimed application in this project we will consider it only.

#### 2.3.1.1.1 Mobile ad hoc networks (MANET):

In fact, MANET is a mobile network which operates without base station infrastructure or centralized administration. Instead of that all nodes cooperate with each other to establish connections while all nodes are constantly changing their position by moving from one side to the other side, hence there should be some protocols to manage these needed random interconnections. These protocols are divided in three groups:

1) Proactive protocols
2) Reactive protocols
3) Hybrid protocols

In proactive protocols to each node a routing table is assigned, and each node forwards its routing table to neighboring nodes to update their routing tables. And the function of these routing tables is to help the data packet to be forwarded to the best possible nodes and reach the destination most reliable and shortest path. In other words, in proactive protocols the shortest paths are the main key in routing. In reactive protocols the source initiates the best route to send the data. Here the commonly used strategy in routing is to flood the network with route request packets. The main disadvantage of this type of protocols is high latency in route finding, it means it takes time for a network to find the best route. Hybrid protocols are a combination of the two previously mentioned protocols (proactive and reactive). For further information in security domain, RSA based encryption in MANETs is used.

# Chapter 3

# LDPC Codes

In Network designing there are two strategies to deal with errors that occurred during transmission, one strategy is to add enough redundant bits to the main message to enable the receiver to detect errors and correct them. In the next strategy we add enough redundant bits to the main message just to make the receiver able to detect errors and maybe in some applications send back a signal showing that the received message was detected in error and ask the source to retransmit it again. Using error detection codes are also referred to as FEC (Forward Error Correction). Error correction codes are various but four well known error correction codes are as follow:

1) Hamming codes.
2) Binary convolutional codes.
3) Reed Solomon codes.
4) Turbo codes.
5) Low-Density Parity check codes

Using error correction techniques in wireless communications will yield a better performance in terms of reliability, it means we will have a better BER in compared to a cooperation without coding techniques. Choosing a suitable error correction code depends on the characteristics of the network in terms data rate, message length, etc. Here in this project, we have used Low-Density parity check codes therefore it seems essential to have a review on it and consider its implemented encoding and decoding steps in our simulation.

## 3.1 LDPC Codes

LDPC codes were firstly proposed by Gallager [17] as his PhD dissertation in 1960s, but according to some reason such as:

1) Very high decoding complexity
2) Introduction of Reed-Solomon codes
3) This common concept that the concatenated Reed-Solomon codes or convolutional codes are good enough for error correction coding

LDPC coding was ignored until 1996 while Mackay discovered that the performance of LDPC codes is as good as turbo codes. They are called Low-Density Parity check because their parity check matrix contains many zeros and just a few number ones, and that is why they are also known as sparse matrices. The applications of LDPC codes vary from digital video broadcasting to WIMAX (IEEE802.16e).

Also, LDPC codes are divided into two classes:

1) **Structured LDPC codes:** the position of ones in parity check matrix is opted to maximize the performance or better to say to optimize the performance of the code.
2) **Random LDPC codes:** the position of ones in parity check matrix is pseudo-randomly positioned.

Structured LDPC codes have less decoding complexity and are more suitable for short lengths codes rather than random LDPCs codes which has better performance for long code lengths. In terms of row weight and column weight a LDPC code is divided into two parts: regular and irregular LDPC codes. In regular LDPC codes row weight and column weight in parity check matrix are fixed and for irregular one they are not.

In fact, a parity check matrix of an LDPC codes can be expressed as a form of bipartite graph named Tanner Graph. The girth is also the shortest cycle in a Tanner Graph and as much as it is a small the LDPC codes will encounter problem in high bit rates. Also a systematic parity check matrix is in the form $H = [I_{n-k} \mid P]$ and then generator matrix will be in the form of: $G = [P^T \mid I_k]$

An example of a parity check matrix and a generator matrix resulted from that:

$$n = 6$$

$$\mathbf{H} = \begin{bmatrix} 1 & 1 & 0 & 1 & 0 & 0 \\ 0 & 1 & 1 & 0 & 1 & 0 \\ 0 & 0 & 1 & 1 & 0 & 1 \end{bmatrix} \quad n - k = 3$$

$$\mathbf{GH^T = 0}$$

$$n = 6$$

$$\mathbf{G} = \left[\begin{array}{ccc:ccc} 0 & 1 & 1 & 1 & 0 & 0 \\ 1 & 1 & 0 & 0 & 1 & 0 \\ 1 & 1 & 1 & 0 & 0 & 1 \end{array}\right] \quad k = 3$$

$$\underbrace{\qquad}_{\mathbf{P}^T} \qquad \underbrace{\qquad}_{\mathbf{I}_k = \mathbf{I}_3}$$

### 3.1.1 LDPC Encoding:

It can be calculated easily from the formula below:

M= Message

G = Generator Matrix

C = Coded Message

$$C = M.\,G \qquad (3\text{-}1)$$

### 3.1.2 LDPC Decoding:

Decoding is done iteratively through belief propagation algorithm or message passing algorithm.

The advantage of decoding in LDPC codes over turbo codes is that here there is no interleaver / deinterleaver, thus LDPC codes do not have the problem of long delays. A decoding is made up of 4 steps: initialization reliability, horizontal step, vertical step and hard decision decoding.

First, we assume the arrival signal is y [i] and parity check matrix is defined as H and generator matrix as G, then: ($\sigma^2$ Rayleigh channel variance)

1) **Initialization Reliability (q):**

The first step to decode the received bits in receiver is initialization which can be done via this formula:

$$2/\sigma^2 \times y[i] \qquad (3\text{-}2)$$

2) **Horizontal Step**:

After initialization there will be a Q matrix as the result of initialization step then in horizontal step the value of each array in the new matrix R are calculated by multiplication of the tangent hyperbolic of other non-zero values of each row of the matrix

$$R_{ij} = \ln\left(\frac{1+\prod tgh\,(qij'/2)}{1-\prod tgh\,(qij'/2)}\right) \qquad j' \in (N_{i/j}) \qquad (3\text{-}3)$$

3) **Vertical Step:**

In vertical matrix the process is like in horizontal step but here instead of multiplication there will be summation and instead of doing operation in a row, the summation will be done in a column.

$$q_{ij} = 2/\sigma^2 \times y[i] + \sum_{i' \in Mj/i} ri'j \qquad (3\text{-}4)$$

4) **Hard Decision Decoding:**

In decoding, decision making is the final step and here in this project a hard decision has been chosen and the output of this step is the main sent message by transmitter even though there will be some bit errors in the extracted message.

$$2/\sigma^2 y_i + \sum_{i \in Mj} rij \qquad (3\text{-}5)$$

## 3.2 Simulation

Here we do a simulation for a direct link communication with two nodes, in one of them no LDPC coding technique is used and in the other one LDPC coding technique is used. The channel is assumed to be a simple AWGN channel. We do this simulation for varying SNRs from 1 to 15 (dB). The parity check matrix of the LDPC is a 100 × 200 matrix with a code rate equal to 0.5. Also, the modulation scheme used here is simple BPSK. As we see, the BER performance of the communication system with LDPC is much better than the one without LDPC.

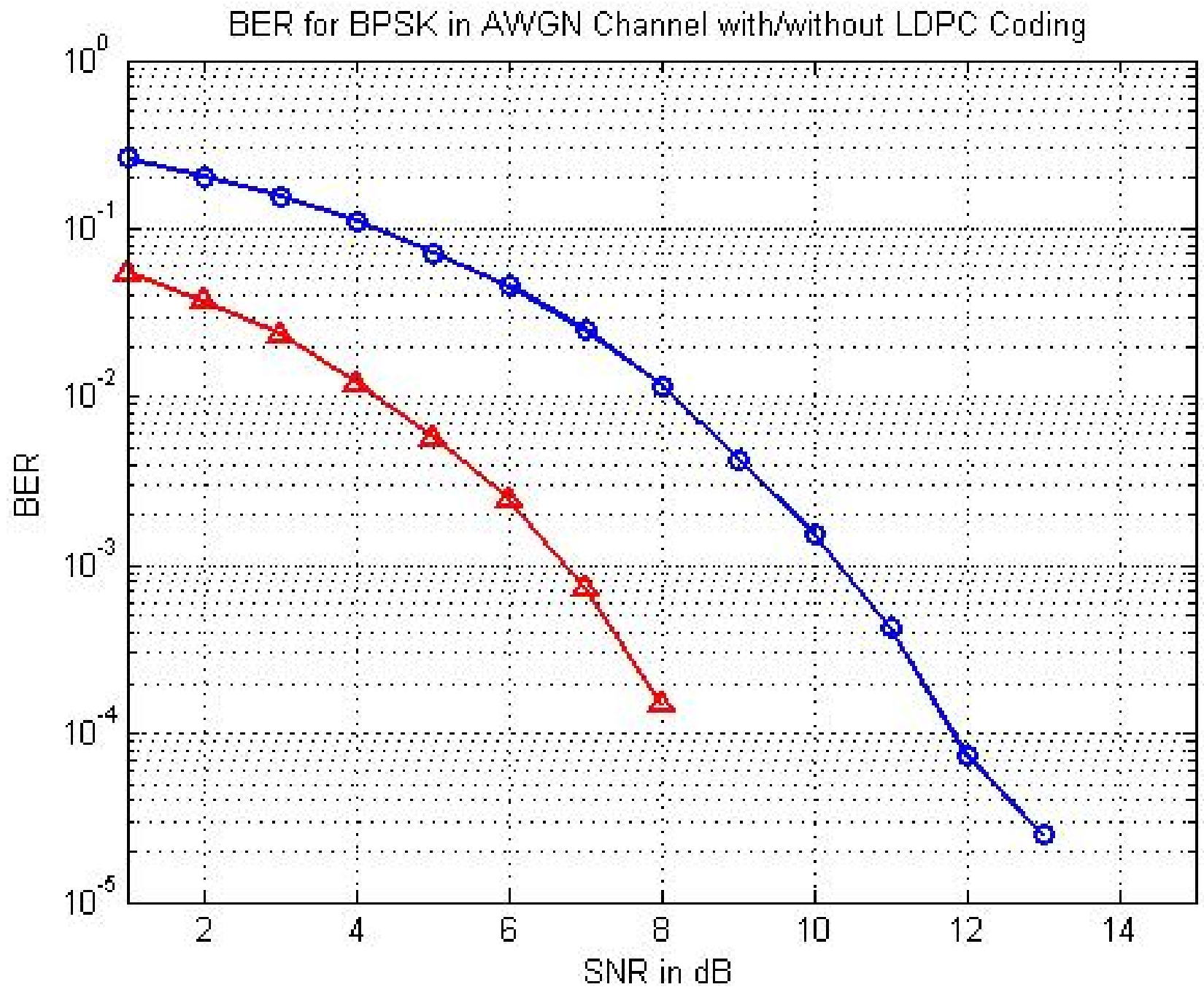

**Figure 9** ----- **With LDPC**

----- **Without LDPC**

# Chapter 4

# LDPC Coding
# In Cooperative
# Communication

In chapter two we discussed cooperative communication and its performance in compared to non-cooperative communication and eventually reached to this conclusion that cooperative communication can improve the performance of a wireless network in terms of bit error rates. In other words by using cooperative communication the reliability of the network is increased and therefore we can have higher data rates in our communications. Also, it means we will have a better spectral efficiency and then we did a simulation in order to compare the performance of these two networks: cooperative and non-cooperative, and the results were as we expected a better BER /SNR performance for the cooperative network. In chapter three we changed the topic and discussed briefly error correction and detection coding techniques and then came to LDPC codes as one of iterative codes with complex decoding process but good performance which can improve bit error rate performance of a communication and we did a simulation on this matter and compared the BER / SNR performance of a direct link communication between two nodes with and without LDPC codes and saw a better performance for LDPC used one. Thus, here we step forward and combine the concepts of chapter two and chapter three by using LDPC codes in a 3- node cooperative network. Therefore, we expect an even better performance of this cooperative network compared to the network without LDPC codes.

## 4.1 Simulation

We consider a 3-node network and assume SNRs of both uplink channels vary from 0 to 15 and the SNR of inter-user channel is constant and equal to 5dB. The other conditions of the network like Rayleigh fading coefficients and length of message are the same as in chapter two. The parity check matrix of the LDPC codes used in this simulation is a 100 × 200 matrix (size) with as mentioned earlier in chapter two mostly zeros. The used LDPC code is a structured one, thus it is more suitable for short length codes. The column weight of the parity check matrix is 2 and row weight is 4 then according to the following formula the code rate of the matrix will be 0.5. (the same parity heck matrix used for simulation of chapter 3)

$$R = 1 - \frac{w_c}{w_r}$$

Therefore, we want to compare the performance of a cooperative communication network (with LDPC) with a cooperative communication network (without LDPC) and with a non-cooperative communication network (better to say a simple one link network without relay).

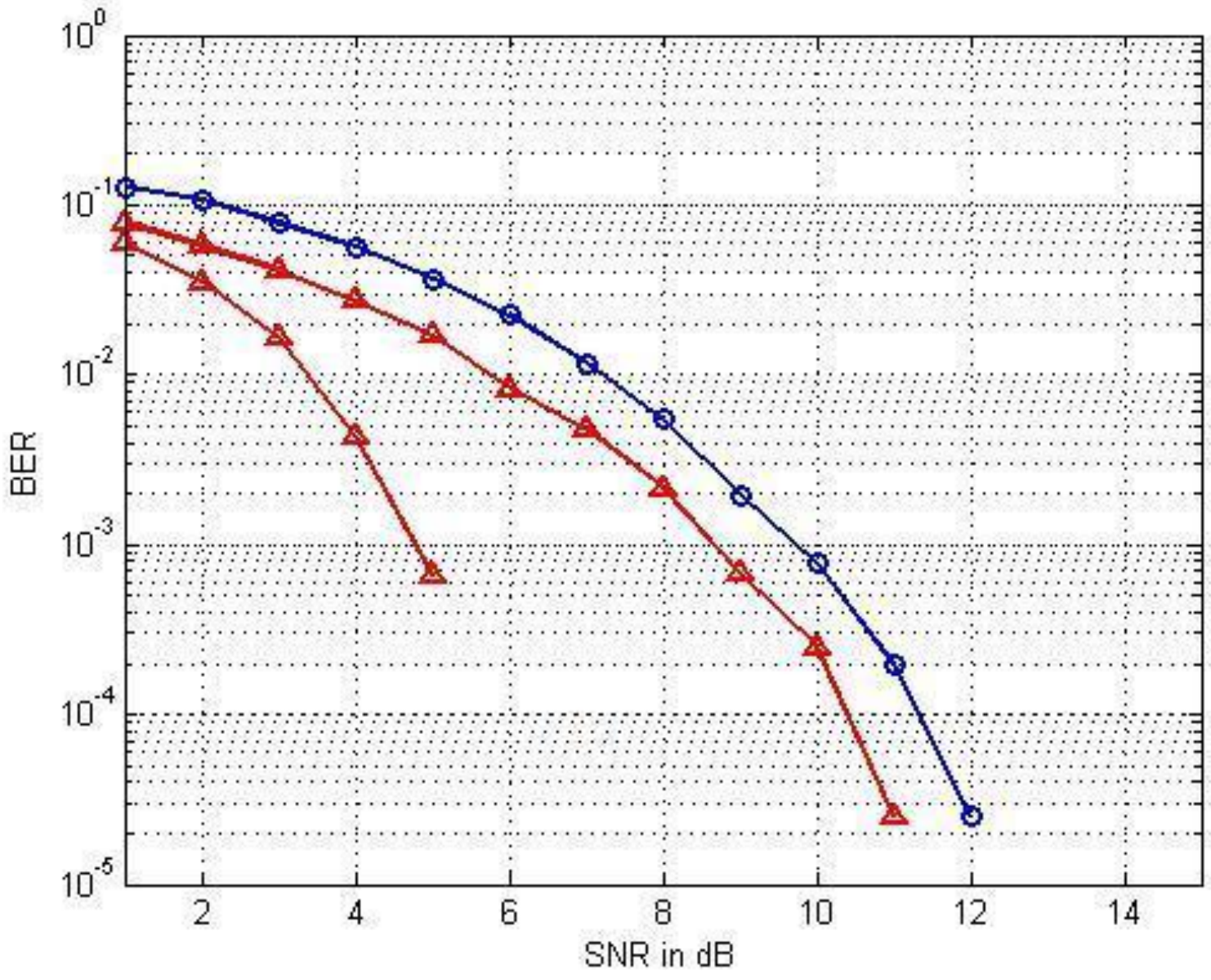

**Figure 10**

**---- Non- Cooperative**

**---- Cooperative Without LDPC (in the middle)**

**---- Cooperative With LDPC ( at the bottom)**

As we can see the performance of a system using cooperative communication is better and more reliable than a system without relay (non-cooperative system). And, when LDPC codes are used in communication the performance becomes even better. Now we do another simulation but only for cooperative network with LDPC codes and the aim of this one is to compare the BER performance of this system for different inter-user channel SNRs (0, 10, 20,100 dB) and the SNR of uplink Rayleigh channels are again varying from 0 to 15. Also notable to know that inter-user channel is an AWGN channel like the simulations in previous chapters The LDPC specifications are all the same as before but here the uplink channels are assumed to be Rayleigh channel and signal power is supposed to be 1(ε=1). Also, the number of data frames

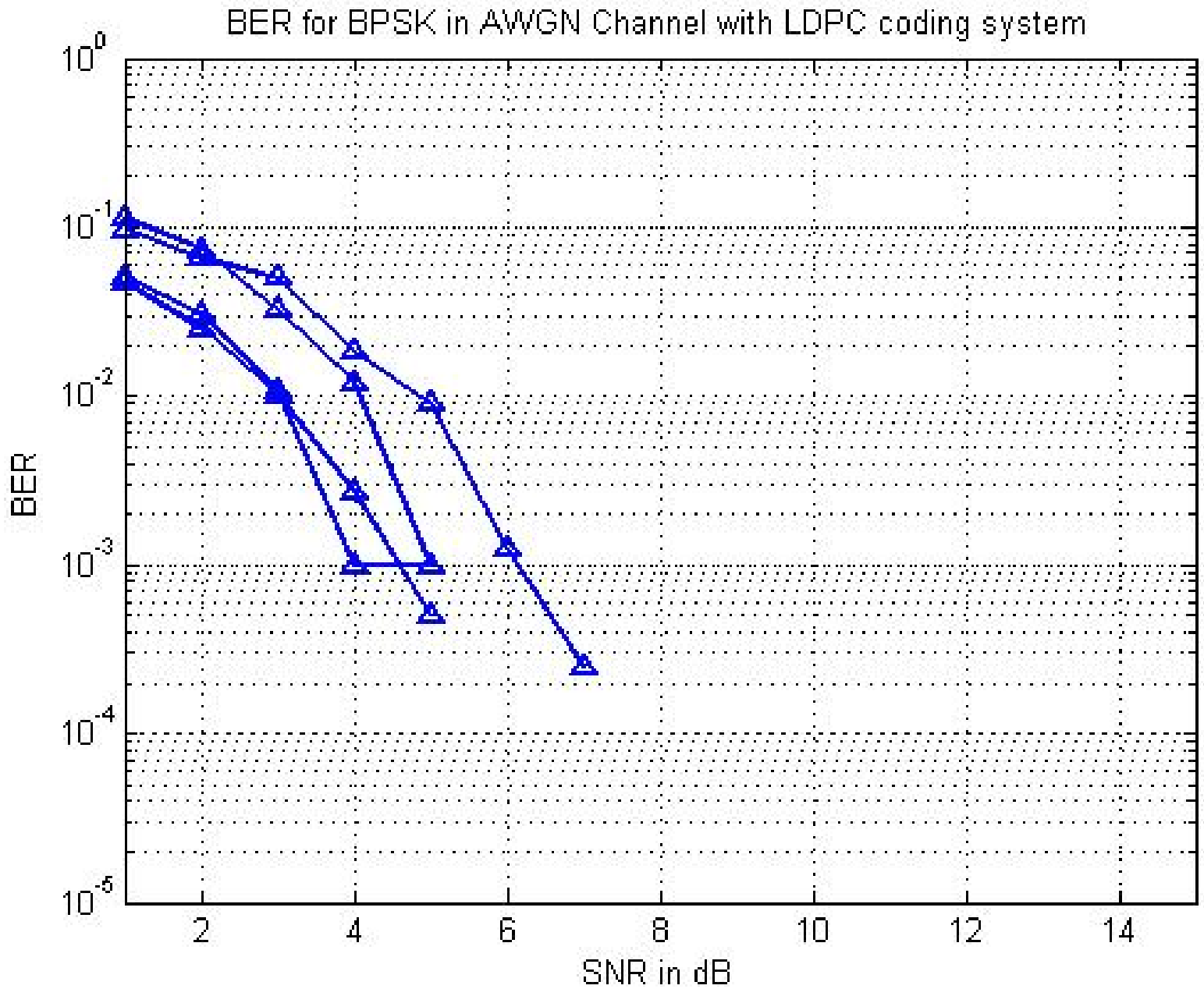

As we see from the to the bottom SNRs are 0, 10, 20 and100. It means as much the level of noise in the inter-user channel is higher the reliability of the message in the output of decoder (in receiver) is lower.

# Chapter 5

# Conclusion

Multipath fading caused by diffraction, refraction, reflection and scattering is a crucial problem for wireless communications and makes it hard for receivers to decide on received signals. Diversity gain is the solution for this problem that provides multiple copies of a sent signal for the receiver to make it able to extract original message of these samples via various techniques like MRC. To provide these multiple samples of a signal for the receiver the method is to use frequency, time or space diversity. To benefit from space diversity, multiple antennas must be used in transmitters and even in receivers. The emergence of MIMO systems was because of this idea. But wireless communication by the time devices is becoming smaller and it creates limits in using multiple antennas in wireless devices. Cooperative communication was proposed to solve this problem by assuming all devices in a wireless network act as partners for the other devices. In fact, each device not only works for itself but also makes a contribution to other devices by relaying their signals to their aimed destinations.

By implementing cooperative communication ideas in wireless networks, we can expect some good results for the network like better BER performance, more reliable communication, higher data rate and less power consumption for a network. In this project we have simulated a network composed of three devices, transmitter, receiver and relay. Uplink channels are assumed to be Rayleigh fading channels and inter-user channel as AWGN channel. Then we simulated the BER performance of the network using cooperative communication and compared it with a non-cooperative communication BER performance where there was no relay in the network. In simulation we saw the BER performance has become better even though there was no error correction technique used in the system. Then we did different simulations for different inter-user channel SNRs and finally realized whatever the SNR is smaller performance of the network becomes worse.

In chapter three we discussed LDPC codes as a type of error correction codes that can improve BER performance of a system and in a simulation of a point-to-point communication with and without LDPC codes, the results were as expected earlier, the performance becomes better. Then in chapter four we implemented LDPC codes in our previous tested cooperative network in chapter two to obtain even better throughput and compare it with the states "cooperative network without LDPC codes" and "non-cooperative network". Overall, the BER performance of cooperative network was better than other states and the BER performance of non-cooperative network was the worst. Finally, we did this simulation for different inter-user channel SNRs as before and the result was quite like that again.

Despite all these advantages of cooperative communication, it still has some important challenges ahead. Demand for great time synchronization and action coordination between users is a big challenge. The second challenge is compromising between system performance and power consumption because every user is also dealing with the other users' transmitted information. Between relaying in this field, decode-and-forward delivers better performance gains than amplify-and-forward.